\documentclass[conference]{IEEEtran}
\IEEEoverridecommandlockouts
\usepackage{cite}
\usepackage{amsmath,amssymb,amsfonts}
\usepackage{graphicx}
\usepackage{textcomp}
\usepackage{xcolor}
\usepackage{flushend}
\usepackage{balance}
\usepackage{epsfig}
\usepackage{epstopdf}
\usepackage[linesnumbered,ruled,vlined]{algorithm2e}
\usepackage{float}
\usepackage{subfig}
\usepackage{times}
\usepackage{url}
\usepackage{verbatim}
\usepackage{bm}
\usepackage{dsfont} 

\newcommand{\one}{\mathds{1}}

\usepackage{tikz}
\usetikzlibrary{arrows.meta,positioning,fit,calc}

\def\BibTeX{{\rm B\kern-.05em{\sc i\kern-.025em b}\kern-.08em
    T\kern-.1667em\lower.7ex\hbox{E}\kern-.125emX}}

\begin{document}

\title{Capacity-Aware Multi-UE Handover Optimization for LEO Satellite Networks Using Dueling DDQN\vspace{-0.1in} 
\thanks{This work was supported in part by the National Science and Technology Council (NSTC) of Taiwan under Grant 113-2926-I-001-502-G and 114-2221-E-003-033.}
}
\author{\IEEEauthorblockN{Po-Heng Chou$^{1,3}$, Chiapin Wang$^{2}$, Chung-Chi Huang$^{2}$, Kuan-Hao Chen$^{2}$, and Chien-Yu Chen$^{2}$}
\IEEEauthorblockA{
$^{1}$Research Center for Information Technology Innovation (CITI), Academia Sinica (AS), Taipei 11529, Taiwan\\
$^{2}$Department of Electrical Engineering, National Taiwan Normal University (NTNU), Taipei 106308 Taiwan\\
$^{3}$Bradley Department of Electrical and Computer Engineering (ECE), Virginia Tech (VT), Alexandria, VA 22305, USA\\
E-mails: d00942015@ntu.edu.tw, chiapin@ntnu.edu.tw, w0905350146@gmail.com,\\ howard25765976@gmail.com, anime4766@gmail.com\vspace{-0.2in}
}
}

\maketitle

\begin{abstract}
Frequent handovers and finite satellite capacity make multi-user
mobility management challenging in low Earth orbit (LEO) satellite
networks. This paper develops a capacity-aware multi-user handover
framework based on a double deep Q-network (DDQN) with a dueling architecture. Both training
and evaluation employ sequential user equipment (UE) processing with
shared satellite occupancy, while action masking excludes invisible or
capacity-infeasible satellites. The state incorporates satellite visibility, achievable
link quality, remaining capacity, previous association, short-term
visibility, and time information. A fixed multi-criteria reward balances throughput, blocking, handover
cost, connection persistence, and short-term visibility, while expert trajectories are used only for replay-buffer initialization. Simulations under different
UE densities and satellite capacities show that the proposed framework
maintains high average throughput per UE and low blocking rates with
moderate handover frequency.
Ablation results show that removing the
remaining-capacity vector has only a limited
effect on performance when sequential occupancy
tracking and feasible-action masking are retained.
\end{abstract}


\vspace{-0.1in}

\section{Introduction}
\label{sec:introduction}

Low Earth orbit (LEO) satellite networks are a key component of
next-generation non-terrestrial networks (NTNs) because they can
provide wide-area, low-latency, and ubiquitous connectivity~\cite{Jamshed2025MCOMSTD}. However, rapid satellite motion
causes frequent changes in visibility and link quality, requiring user
equipment (UE) to perform repeated handovers. Under finite satellite
capacity, these handovers are further coupled across UEs because
multiple users may compete for the same satellite resources. Therefore,
handover decisions must jointly consider link quality, service
availability, resource occupancy, and connection stability.

Existing LEO handover methods include rule-based~\cite{Papapetrou2004IJSCN}, optimization-based~\cite{Wu2016CL,Kang2024ICC,Huang2025ICCC,Leyva2023ICC}, and learning-based approaches~\cite{Badini2023ICC,Yang2024TVT,Zhao2026TWC,Zhang2026TVT}.
Rule-based methods are computationally simple but often rely on
instantaneous criteria such as visibility time or available resources.
Optimization-based methods can explicitly incorporate capacity and
handover costs, but may require repeated combinatorial optimization or
future network information. Deep reinforcement learning (DRL) provides
an alternative by learning long-term satellite-selection policies from
network interactions and has also been applied to other LEO network
functions, such as beam positioning~\cite{Chou2026ICCWkshps}.
Recent studies further show the importance of multi-objective handover
and resource-aware association in dense LEO networks
\cite{Sun2024CL,Sun2026TAES,Dou2025TCOM,Yang2026TWC}.

A key difficulty in multi-UE handover is that satellite assignments
cannot be optimized independently. Once one UE occupies a satellite,
the remaining capacity available to subsequently processed UEs changes.
Ignoring this interaction can produce infeasible or overloaded
associations. Moreover, when several visible satellites have similar
conditions, conventional value-based learning may have difficulty
distinguishing their relative actions. These observations motivate a
capacity-aware sequential decision framework together with a dueling
network architecture \cite{Mnih2015DQN,Hasselt2016,Wang2016Dueling}.

In this paper, we develop a capacity-aware multi-UE
handover framework based on a double deep Q-network
(DDQN) with a dueling architecture. Both training and
evaluation process UEs sequentially within each time slot using a
shared satellite-occupancy state. After every UE decision, the
occupancy is immediately updated, and invisible or capacity-infeasible
satellites are removed through action masking. The state further
includes normalized achievable rate, remaining capacity, previous
association, one-slot-ahead visibility, and normalized time. A fixed
multi-criteria reward balances average throughput per UE, blocking, handover
cost, connection persistence, and short-term visibility. Expert trajectories generated by minimum satellite handover for bandwidth
oversubscription (MSHBO)~\cite{Kang2024ICC} are used only to
initialize the replay buffer and do not constrain the learned policy.

The main contributions are summarized as follows:
\begin{itemize}
    \item We formulate multi-UE LEO handover as a
    capacity-coupled sequential decision problem that
    jointly optimizes throughput, blocking, and handover
    cost under time-varying visibility and finite capacity.

    \item We develop a capacity-aware Dueling DDQN
    framework integrating shared occupancy tracking,
    capacity-aware states, sequential UE processing,
    and feasible-action masking to prevent
    infeasible associations.

    \item We evaluate the framework against learning-based
    and heuristic baselines under varying UE densities
    and satellite capacities, with ablation studies
    and publicly available source code.
\end{itemize}

\section{System Model}
\label{sec:system_model}

\subsection{Network Model}

We consider a LEO satellite network consisting of $S$ satellites
and $U$ ground UEs. The satellite and UE sets are denoted by
$\mathcal{S}=\{1,\ldots,S\}$ and $\mathcal{U}=\{1,\ldots,U\}$,
respectively. The observation interval is divided into $T$ discrete
time slots indexed by $t\in\mathcal{T}=\{1,\ldots,T\}$ to capture
the time-varying satellite visibility, link quality, and resource
occupancy.

\begin{figure}[t]
\centering
\includegraphics[width=\linewidth]{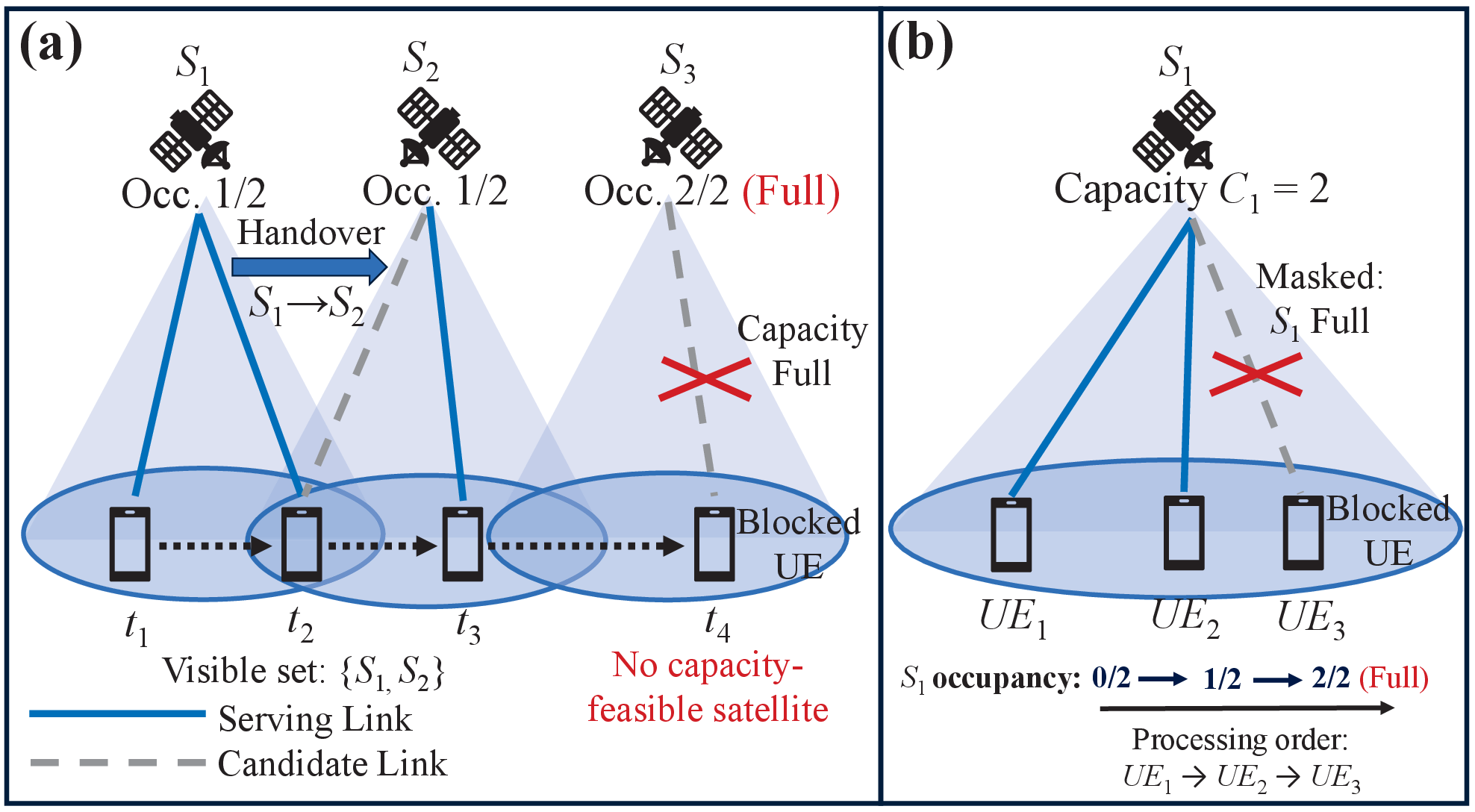}
\caption{Illustration of the considered LEO handover scenarios:
(a) time-varying satellite visibility, handover, and blocking;
(b) capacity-coupled sequential multi-UE association.}
\vspace{-0.1in}
\label{fig:system_model}
\end{figure}

As illustrated in Fig.~\ref{fig:system_model}(a), a UE may be
simultaneously covered by multiple LEO satellites because of overlapping
satellite footprints, and its serving satellite may change over time as
the visibility conditions evolve. Let $v_{u,s}(t)\in\{0,1\}$ denote
the visibility indicator between UE $u$ and satellite $s$ at time slot
$t$. Specifically, $v_{u,s}(t)=1$ if satellite $s$ satisfies the
minimum elevation-angle requirement of UE $u$, and $v_{u,s}(t)=0$
otherwise. Accordingly, the visible satellite set of UE $u$ is
$\mathcal{S}_u(t)=\{s\in\mathcal{S}\mid v_{u,s}(t)=1\}$.

Let $x_{u,s}(t)\in\{0,1\}$ denote the association decision, where
$x_{u,s}(t)=1$ indicates that UE $u$ is served by satellite $s$ at
time slot $t$. Since an association is permitted only when the satellite is visible,
$x_{u,s}(t)\leq v_{u,s}(t)$ for all $u$, $s$, and $t$.

To explicitly represent service interruption, let
$b_u(t)\in\{0,1\}$ denote the blocking indicator of UE $u$.
Each UE is either associated with one visible satellite or placed in
the blocking state, i.e.,
$\sum_{s\in\mathcal{S}_u(t)}x_{u,s}(t)+b_u(t)=1$.

Each satellite has a finite service capacity.
Let $C_s$ denote the maximum number of UEs
that satellite $s$ can simultaneously serve,
and let $L_s(t)$ denote its running occupancy,
which is updated after each UE decision
within time slot $t$. After all UE
decisions in slot $t$ are completed,
$L_s(t)=\sum_{u\in\mathcal{U}}x_{u,s}(t)$.
The satellite-capacity constraint is $L_s(t)\leq C_s$ for all $s$
and $t$.

As illustrated in Fig.~\ref{fig:system_model}(b), UE association
decisions are capacity-coupled. Within a time slot, multiple UEs may
contend for the limited capacity of the same visible satellite. Once a
UE is associated with that satellite, its occupancy is immediately
updated, which changes the remaining capacity available to subsequently
processed UEs. Therefore, a later UE may be blocked even when the
satellite remains visible. The corresponding capacity-aware state
representation and action-masking mechanism are introduced in
Sec.~\ref{sec:proposed_method}.

\subsection{Channel and Rate Model}

The achievable rate between UE $u$ and satellite $s$ at time slot $t$
is modeled as
\begin{equation}
R_{u,s}(t)
=
B\mathbb{E}_{g}
\left[
\log_2
\left(
1+g_{u,s}(t)\gamma_{u,s}(t)
\right)
\right],
\label{eq:rate_model}
\end{equation}
where $B$ is the per-link bandwidth, $\gamma_{u,s}(t)$ is the average
signal-to-noise ratio (SNR) before small-scale fading,
$g_{u,s}(t)$ is the normalized fading power gain, and
$\mathbb{E}_{g}[\cdot]$ denotes expectation over the fading
distribution.

The rate generator uses $B=500$ MHz, a carrier frequency of $28$ GHz,
an effective isotropic radiated power (EIRP) of $50$ dBW, a receiver
gain of $40$ dBi, a system noise temperature of $400$ K, an additional
link loss of $3$ dB, and a reference satellite altitude of $1300$ km.
Rates are SNR-based and neglect inter-user, inter-beam, and
inter-satellite interference. Shadowed-Rician fading is generated using a Rice--Gamma approximation
with Rician $K$-factor $10$ dB, shadowing shape parameter $m=1.5$,
normalized channel-power parameter $\Omega=1$, and 200 Monte Carlo
samples. The bandwidth is
fixed per UE and per link; hence, $C_s$ limits the number of admitted
UEs rather than dividing the available bandwidth. The resulting
achievable-rate table in Mbps is shared by all schemes.

The average throughput per UE is defined as
\begin{equation}
\overline{R}
=
\frac{1}{UT}
\sum_{t=1}^{T}
\sum_{u\in\mathcal{U}}
\sum_{s\in\mathcal{S}_u(t)}
x_{u,s}(t)R_{u,s}(t).
\label{eq:average_throughput}
\end{equation}
A blocked UE contributes zero throughput. Raw rates in Mbps are used
for evaluation, whereas normalized rates are used for state and reward
construction.

\subsection{Handover Model}
Let $\sigma_u(t)$ denote the serving satellite of UE $u$ at time slot
$t$, with $\sigma_u(t)=\varnothing$ when the UE is blocked. A satellite
handover occurs when the UE changes from one serving satellite to
another while remaining connected. Accordingly,
$h_u(t)=\one\{\sigma_u(t-1)\neq\varnothing,\,
\sigma_u(t)\neq\varnothing,\,
\sigma_u(t)\neq\sigma_u(t-1)\}$ for $t\geq2$, where
$\one\{\cdot\}$ denotes the indicator function.
For initialization, we set $\sigma_u(0)=\varnothing$ and $h_u(1)=0$.

The average number of handovers per UE over the observation interval is
$\mathcal{H}
=
\frac{1}{U}
\sum_{u\in\mathcal{U}}
\sum_{t=2}^{T}
h_u(t)$.
Transitions into or out of the blocking state are not counted as
successful satellite-to-satellite handovers because they correspond to
service interruption or reconnection.

\subsection{Blocking Model}

Limited satellite resources may prevent a UE from being admitted even
when one or more satellites remain visible. Blocking denotes an
unserved UE state at a given time slot. Let $b_u(t)\in\{0,1\}$
indicate whether UE $u$ is not associated with any satellite at time
slot $t$.
The average blocking rate is defined as
$\mathcal{B}
=
\frac{1}{UT}
\sum_{t=1}^{T}
\sum_{u\in\mathcal{U}}
b_u(t)$, and is reported as a percentage.

\section{Problem Formulation}
\label{sec:problem_formulation}

Based on the system model described in Sec.~\ref{sec:system_model}, the
handover problem jointly considers average-throughput-per-UE
maximization, blocking-rate reduction, and handover-cost
minimization under satellite visibility and service-capacity
constraints. The resulting optimization
problem is formulated as
\begin{subequations}
\begin{align}
\max_{\{x_{u,s}(t),b_u(t)\}}
\quad &
\overline{R}
-
w_{\mathrm{b}}\mathcal{B}
-
w_{\mathrm{h}}\mathcal{H}
\label{eq:obj}
\\
\text{s.t.}
\quad &
\sum_{s\in\mathcal{S}_u(t)}
x_{u,s}(t)
+
b_u(t)
=
1,
\quad
\forall u,t,
\label{eq:const1}
\\
&
x_{u,s}(t)
\leq
v_{u,s}(t),
\quad
\forall u,s,t,
\label{eq:const2}
\\
&
\sum_{u\in\mathcal{U}}
x_{u,s}(t)
\leq
C_s,
\quad
\forall s,t,
\label{eq:const3}
\\
&
x_{u,s}(t)\in\{0,1\},
\quad
b_u(t)\in\{0,1\},
\quad
\forall u,s,t.
\label{eq:const4}
\end{align}
\label{eq:optimization_problem}
\end{subequations}
\hspace{-0.07in}where $w_{\mathrm{b}}\geq0$ and $w_{\mathrm{h}}\geq0$
control the relative importance of blocking and handover cost.
Constraint~(\ref{eq:const1}) requires each UE to be either associated
with one visible satellite or blocked,
(\ref{eq:const2}) enforces visibility,
(\ref{eq:const3}) limits satellite capacity, and
(\ref{eq:const4}) specifies the binary variables.

The decisions are coupled across UEs because each association changes
the capacity available to subsequently processed UEs, and across time
because the current serving satellite affects future handovers and
service availability. Repeatedly solving
(\ref{eq:optimization_problem}) is therefore combinatorial. We instead
formulate the handover process as a Markov decision process (MDP) and
learn a capacity-aware Dueling DDQN policy using link-quality,
remaining-capacity, association, visibility, and time information.
Action masking and sequential occupancy updates enforce capacity
feasibility during both training and evaluation.

The DRL reward introduced in Sec.~\ref{sec:proposed_method} is a shaped
surrogate of (\ref{eq:obj}); besides throughput, blocking, and handover
terms, it includes persistence and short-term visibility terms to
facilitate policy learning.

\section{Proposed Capacity-Aware Dueling DDQN Handover Framework}
\label{sec:proposed_method}

In this section, the capacity-coupled multi-UE handover problem is
modeled as an MDP, and a Dueling DDQN framework is developed to learn
long-term satellite-selection policies.
The proposed framework incorporates satellite visibility, achievable
link quality, remaining service capacity, previous association, and
short-term visibility information. Capacity feasibility is maintained
through connection-persistent occupancy tracking, sequential UE
processing, and action masking.

\subsection{Capacity-Aware MDP Formulation}

The handover process is modeled as an MDP
$\mathcal{M}=(\mathcal{X},\mathcal{A},\mathcal{P},\mathcal{R},\delta)$,
where $\mathcal{X}$, $\mathcal{A}$, $\mathcal{P}$, $\mathcal{R}$,
and $\delta$ denote the state space, action space, transition
probability, reward function, and discount factor, respectively.

For UE $u$ at time slot $t$, the state is defined as
\begin{equation}
\bm{s}_u(t)
=
[
\bm{v}_u(t),
\bm{\rho}(t),
\bm{\ell}_u(t),
\bm{v}_u(t+1),
\tilde{\bm{r}}_u(t),
\tau(t)
],
\label{eq:state}
\end{equation}
where $\bm{v}_u(t)\in\{0,1\}^{S}$ is the current satellite-visibility
vector, $\bm{\rho}(t)=[\rho_1(t),\ldots,\rho_S(t)]$ is the
remaining-capacity-ratio vector,
$\bm{\ell}_u(t)\in\{0,1\}^{S}$ is the one-hot representation of the
current retained serving satellite before the decision,
$\bm{v}_u(t+1)$ provides one-slot-ahead visibility information,
$\tilde{\bm{r}}_u(t)\in[0,1]^S$ is the normalized achievable-rate
vector, and $\tau(t)=t/T$ is the normalized time index. When the
UE has no retained serving satellite, $\bm{\ell}_u(t)$ is the all-zero
vector. For the terminal slot, we define
$\bm{v}_u(T+1)=\bm{v}_u(T)$, consistent with the boundary handling of
the one-slot-ahead visibility information. The resulting state
dimension is $5S+1$.

The normalized achievable rate is defined as
$\tilde{R}_{u,s}(t)=R_{u,s}(t)/R_{\max}$, where
$R_{\max}=2946.289$ Mbps is the maximum achievable rate in the adopted
rate table, and
$\tilde{\bm{r}}_u(t)
=[\tilde{R}_{u,1}(t),\ldots,\tilde{R}_{u,S}(t)]$.
Normalization prevents the Mbps-scale rate values from dominating the
other state and reward components.

The remaining-capacity ratio is
$\rho_s(t)=\max\{0,C_s-L_s(t)\}/C_s$, where $L_s(t)$ denotes the
current satellite occupancy. Active associations are carried across
consecutive time slots. A capacity reservation is released when the
serving satellite becomes invisible, the UE enters the blocking state,
or a successful handover occurs. Within a time slot, $L_s(t)$ is
updated immediately after each UE decision.

The complete action space is
$\mathcal{A}=\mathcal{S}\cup\{a^{\mathrm{blk}}\}$, where
$a^{\mathrm{blk}}$ denotes the blocking action. A visible satellite is
legal when it has remaining capacity. The currently retained serving
satellite remains legal even when its occupancy equals $C_s$, because
maintaining an existing association does not consume an additional
capacity unit. The blocking action is always legal.

To reduce the candidate action set, a Top-$K_{\mathrm{cand}}$ rule
with $K_{\mathrm{cand}}=3$ is used.
If the current serving satellite remains legal, it is retained in the
candidate set, and the remaining $K_{\mathrm{cand}}-1$ positions are filled by legal
satellites with the largest remaining capacities. If no serving satellite is retained, the $K_{\mathrm{cand}}$ legal
satellites with the largest remaining capacities are selected. When
no more than $K_{\mathrm{cand}}$ legal satellites are available, all
of them are retained. The blocking action
is always included. Actions outside this candidate set are masked before action selection.
Let $\widehat{\mathcal{A}}_u(t)\subseteq\mathcal{A}$ denote the
resulting masked candidate action set, and let
$a_u(t)\in\widehat{\mathcal{A}}_u(t)$ denote the action selected
for UE $u$ at time slot $t$.

To capture the coupling among multiple UEs, both training and
evaluation process UEs sequentially within each time slot. The UE
processing order is randomly permuted without replacement at every
slot. All UEs share the same satellite-occupancy state, and each
association decision immediately changes the capacity observed by
subsequently processed UEs. Separate seeded random-order streams are
used for training, expert-trajectory replay, and evaluation.

During training, masked $\epsilon$-greedy exploration is performed over
the Top-$K_{\mathrm{cand}}$ candidate set. In addition, when the previously serving
satellite remains legal, the training behavior policy retains that
satellite with probability $p_{\mathrm{stick}}=0.2$. This persistence
mechanism is used only during training. Evaluation uses greedy action
selection over the same Top-$K_{\mathrm{cand}}$ candidate set without the stochastic
sticky override.

The transition dynamics depend on satellite motion, visibility
evolution, link quality, retained associations, and the shared
capacity state, and are learned implicitly through interactions with
the environment. For compactness, $\bm{s}_n$, $a_n$, $r_n$, and
$\widehat{\mathcal{A}}_n$ denote the state,
selected action, reward, and masked candidate
action set at generic decision step $n$,
respectively.

\subsection{Fixed Multi-Criteria Reward}

To learn a long-term handover policy, the instantaneous reward balances
service availability, achievable link quality, handover cost, connection
persistence, and short-term satellite visibility.
Let $K_{\mathrm{look}}$ denote the maximum look-ahead horizon and
$K_t=\min\{K_{\mathrm{look}},T-t\}$ its effective value at time slot
$t$.
For UE $u$ at time
slot $t$, the reward is defined as
\begin{align}
r_u(t)
={}&-\lambda_{\mathrm b}b_u(t)
-\lambda_{\mathrm h}h_u(t)
+\lambda_{\mathrm t}\tilde{R}_{u,a_u(t)}(t)
\nonumber\\
&+b_{\mathrm{stay}}
\one\{\sigma_u(t)=\sigma_u(t-1)\neq\varnothing\}
\nonumber\\
&-\eta\one\{a_u(t)\in\mathcal S\}
\sum_{k=1}^{K_t}
\one\{v_{u,a_u(t)}(t+k)=0\}.
\label{eq:reward}
\end{align}
where $\tilde{R}_{u,a_u(t)}(t)$ denotes the normalized achievable rate
of the selected satellite. For the blocking action,
$\tilde{R}_{u,a^{\mathrm{blk}}}(t)=0$.
The summation is zero when $K_t=0$.
For the blocking action, the future-visibility
penalty is defined as zero.

The first term penalizes service blocking, while the second term
penalizes satellite-to-satellite handovers. The third term encourages
selection of satellites providing high achievable rates. Unlike the
raw rate in Mbps used for performance evaluation, the normalized rate
is used in the reward to maintain a comparable numerical scale among
the reward components.

The fourth term provides a small connection-persistence reward when a
UE remains associated with the same satellite across consecutive time
slots. The final term penalizes the selection of a satellite that
becomes invisible within the next $K_{\mathrm{look}}$ slots, thereby
discouraging associations with insufficient short-term visibility.

Here, $\lambda_{\mathrm b}$, $\lambda_{\mathrm h}$, and
$\lambda_{\mathrm t}$ denote the blocking, handover, and throughput
weights, respectively; $b_{\mathrm{stay}}$ is the persistence reward,
and $\eta$ is the future-visibility penalty weight. We set
$(\lambda_{\mathrm b},\lambda_{\mathrm h},\lambda_{\mathrm t})
=(12,3,2)$, $b_{\mathrm{stay}}=0.05$, $K_{\mathrm{look}}=5$, and
$\eta=0.1$.
These coefficients remain unchanged throughout training.

As discussed in Sec.~\ref{sec:problem_formulation}, the reward in
(\ref{eq:reward}) is a shaped surrogate of the system-level objective
rather than an exact reformulation. The blocking, handover, and
throughput-related terms correspond directly to the primary performance
criteria, whereas the persistence and short-term-visibility terms are
introduced to facilitate stable sequential decision making.

\begin{figure*}[!t]
\centering
\includegraphics[width=.79\textwidth]
{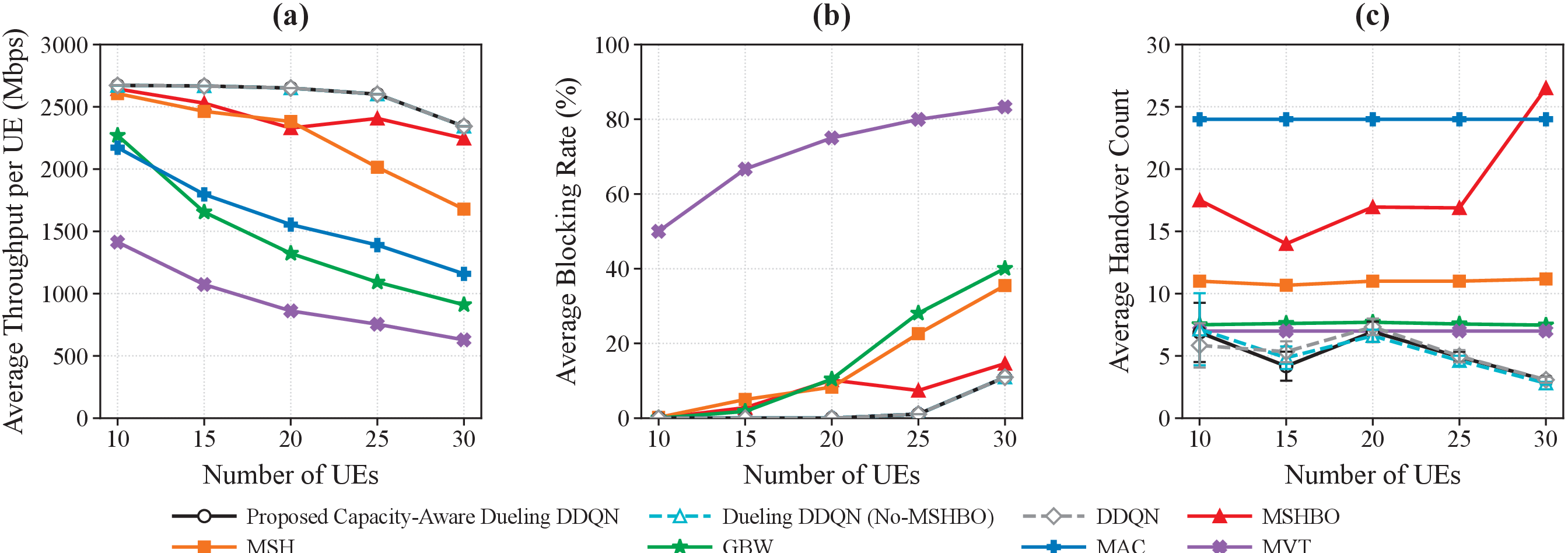}
\caption{Performance versus the number of UEs with
satellite capacity $C_s=5$: (a) average throughput
per UE, (b) average blocking rate, and (c) average
number of handovers per UE.}
\label{fig:ue_sweep}
\vspace{-0.2in}
\end{figure*}

\begin{figure*}[!t]
\centering
\includegraphics[width=.79\textwidth]
{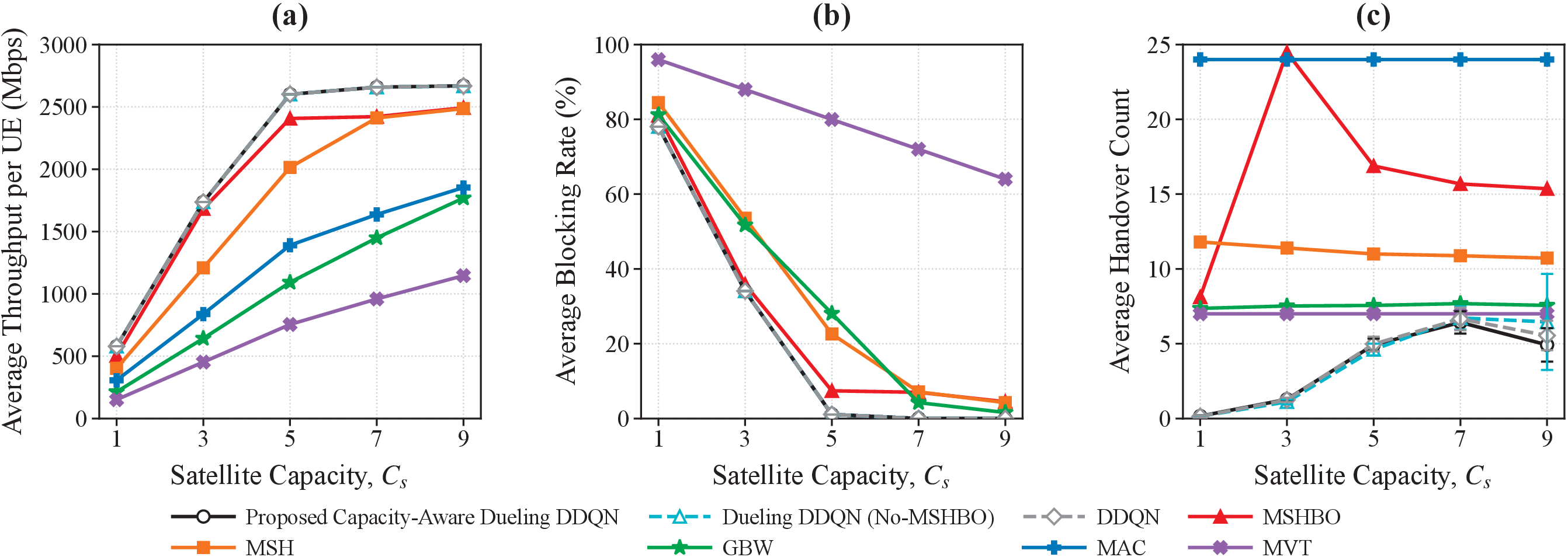}
\caption{Capacity sensitivity with $U=25$: (a) average throughput per
UE, (b) average blocking rate, and (c) average number of handovers per
UE.}
\label{fig:capacity_sensitivity}
\vspace{-0.3in}
\end{figure*}

\subsection{Dueling DDQN Learning}

A DDQN~\cite{Hasselt2016} with a dueling network
architecture~\cite{Wang2016Dueling} is employed to learn the
capacity-aware handover policy. The online and target networks are
parameterized by $\boldsymbol{\theta}$ and
$\boldsymbol{\theta}^{-}$, respectively. The dueling architecture
decomposes the action-value function as
\begin{equation}
\begin{aligned}
Q(\bm{s},a;\boldsymbol{\theta})
={}&V(\bm{s};\boldsymbol{\theta}_{v})
+A(\bm{s},a;\boldsymbol{\theta}_{a})\\
&-\frac{1}{|\mathcal{A}|}
\sum_{a'\in\mathcal{A}}
A(\bm{s},a';\boldsymbol{\theta}_{a}),
\end{aligned}
\label{eq:dueling_q}
\end{equation}
where $V(\bm{s};\boldsymbol{\theta}_{v})$ and
$A(\bm{s},a;\boldsymbol{\theta}_{a})$ denote
the state-value and action-advantage functions,
respectively, with $\boldsymbol{\theta}_{v}$
and $\boldsymbol{\theta}_{a}$ representing
their corresponding network parameters.

For a sampled transition
$(\bm{s}_n,a_n,r_n,\bm{s}_{n+1},d_n)$, let
$a_{n+1}^{\star}
=\arg\max_{a'\in\widehat{\mathcal{A}}_{n+1}}
Q(\bm{s}_{n+1},a';\boldsymbol{\theta})$
denote the greedy action selected by the online network.
The DDQN target is then
\begin{equation}
y_n
=
r_n+(1-d_n)\delta
Q(\bm{s}_{n+1},a_{n+1}^{\star};
\boldsymbol{\theta}^{-}),
\label{eq:ddqn_target}
\end{equation}
where $d_n\in\{0,1\}$ is the terminal-state
indicator and $\widehat{\mathcal{A}}_{n+1}$
is the masked Top-$K_{\mathrm{cand}}$
candidate action set. The online network
selects $a_{n+1}^{\star}$, whereas the target
network evaluates that action, thereby reducing
overestimation bias. The online network is
trained by minimizing the Huber loss between
$y_n$ and $Q(\bm{s}_n,a_n;\boldsymbol{\theta})$.
Experience replay, gradient clipping, and
periodic target-network updates are adopted
to improve training stability~\cite{Mnih2015DQN,Hasselt2016}.

During each training slot, the state and candidate action set of each
UE are constructed using the latest shared occupancy after all
preceding UE decisions in that slot. The resulting experience is
stored in the replay buffer, thereby exposing the agent to the same
capacity-coupled multi-UE interaction encountered during evaluation.

To initialize the replay buffer, MSHBO trajectories~\cite{Kang2024ICC}
are replayed using the same connection-persistent capacity environment
and an independently shuffled UE order. Expert satellite actions that
are legal under the full visibility-and-capacity mask, together with
explicit blocking actions, are retained as demonstration transitions.
The online network is pretrained using these transitions before
standard agent interaction. During subsequent training, the replay
buffer contains both demonstration and agent-generated transitions.
The demonstrations provide initialization only and do not constrain
the final learned policy.

\section{Simulation Results}

\subsection{Simulation Setup}
\label{sec:simulation_setup}

The simulation data are generated from a Telesat Lightspeed LEO
constellation using Ansys Systems Tool Kit (STK)~\cite{AnsysSTK}.%
\footnote{Source code and configuration files are available at:
\texttt{github.com/BoneZhou/Capacity-Aware-LEO-Handover-DDQN}.}
After visibility-based preprocessing, 50 candidate satellites are
retained. The minimum elevation angle is $20^\circ$, and the evaluation
interval contains $T=3599$ one-second time slots. We vary
$U\in\{10,15,20,25,30\}$ with $C_s=5$ and evaluate capacity sensitivity
using $U=25$ and $C_s\in\{1,3,5,7,9\}$.
All candidate satellites have identical
service capacities within each experiment.
Each learning-based method is independently trained
using five random seeds with identical hyperparameters.
All trained models are evaluated using the same fixed
evaluation seed. For each experimental setting,
results are averaged across the five independently
trained models, with error bars indicating one
sample standard deviation.

The proposed Dueling DDQN uses two hidden layers with 512 and 256
neurons and is trained for 300 episodes using Adam with learning rate $10^{-3}$ and discount factor
$\delta=0.99$. The replay-buffer size and
mini-batch size are 200,000 and 256, respectively; the target network
is updated every 2000 global time slots. The exploration probability $\epsilon$ decays exponentially from
$1.0$ to $0.05$ with decay constant 30,000,
and gradient clipping with maximum norm 1 is applied. MSHBO expert
trajectories initialize the replay buffer, followed by 2000 pretraining
updates. We set $K_{\mathrm{cand}}=3$, $p_{\mathrm{stick}}=0.2$,
$(\lambda_{\mathrm b},\lambda_{\mathrm h},\lambda_{\mathrm t})
=(12,3,2)$, $b_{\mathrm{stay}}=0.05$, $K_{\mathrm{look}}=5$, and
$\eta=0.1$.

The proposed method is compared with Dueling DDQN without
MSHBO initialization, standard DDQN~\cite{Hasselt2016},
minimum satellite handover (MSH) and
MSHBO~\cite{Kang2024ICC}, graph-based weighting
(GBW)~\cite{Hozayen2022GCWkshps}, and maximum
available channels (MAC) and maximum visible
time (MVT)~\cite{Papapetrou2004IJSCN}.
All learning-based methods use the same simulation
environment, capacity-accounting rules, reward parameters,
and evaluation procedure unless otherwise specified.

\begin{figure*}[!t]
\centering
\includegraphics[width=.79\textwidth]
{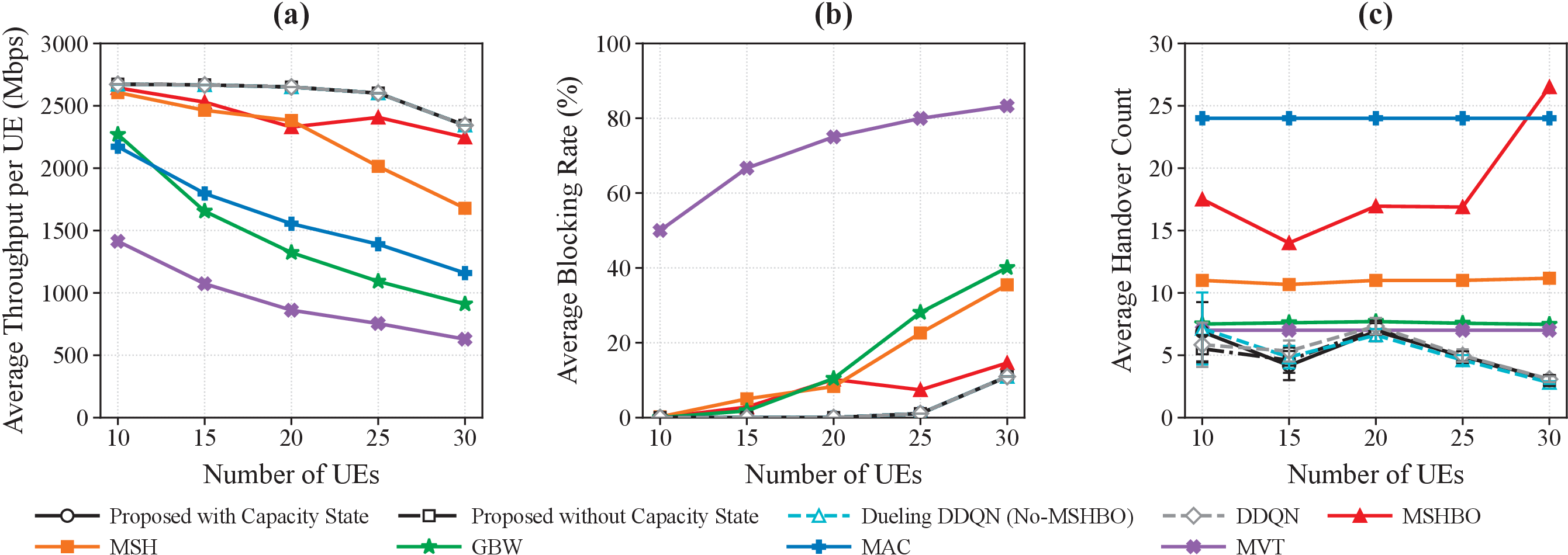}
\caption{Capacity-state ablation under different numbers of UEs with
$C_s=5$: (a) average throughput per UE, (b) average blocking rate,
and (c) average number of handovers per UE.}
\label{fig:capacity_ablation}
\vspace{-0.3in}
\end{figure*}

\subsection{Performance Evaluation}

Fig.~\ref{fig:ue_sweep} compares the schemes as the number of UEs
increases with $C_s=5$. Average throughput per UE decreases with UE
density because more users compete for finite satellite capacity.
The learning-based variants maintain higher throughput and lower
blocking than the heuristic baselines. The proposed method maintains
moderate handover frequency, while Dueling DDQN without MSHBO generally
yields fewer handovers, indicating that expert initialization changes
the operating point without consistently improving final performance.

For $U=25$ and $C_s=5$, the proposed method
achieves $2601.162 \pm 0.407$ Mbps throughput
per UE, a blocking rate of
$1.0803 \pm 0.0030\%$, and
$4.920 \pm 0.420$ handovers per UE
across five independent training seeds.
The small throughput and blocking-rate
variations indicate stable performance
across independent training runs under
this setting.
The other three learning-based variants exhibit
similarly small throughput variations, with
standard deviations ranging from
$0.176$ to $0.323$ Mbps.

\subsection{Capacity Sensitivity}

Fig.~\ref{fig:capacity_sensitivity} evaluates different
satellite capacities with $U=25$. As $C_s$ increases,
throughput per UE improves and blocking decreases.
The throughput gains gradually diminish as satellite
capacity becomes sufficient to accommodate most UEs.
The learning-based methods achieve near-zero blocking
at moderate capacities while generally requiring fewer
handovers than MSHBO and MAC. Their handover counts
may increase modestly as fewer UEs remain blocked,
allowing more continuous connections to undergo
satellite-to-satellite handovers.

\subsection{Capacity-Aware Ablation Study}

Fig.~\ref{fig:capacity_ablation} compares the proposed
method with and without the remaining-capacity vector
$\bm{\rho}(t)$. Removing this vector produces only
minor changes in throughput, blocking, and handover
performance, suggesting that sequential occupancy
updates and feasible-action masking retain much of
the capacity-aware behavior.

\section{Conclusion}

This paper developed a capacity-aware multi-UE handover framework for
LEO satellite networks using Dueling DDQN. Shared occupancy tracking,
sequential UE processing, feasible-action masking, and capacity-related
state information jointly capture multi-UE coupling under finite
satellite capacity. Simulations across different UE densities and
satellite capacities demonstrate high average throughput per UE, low
blocking, and moderate handover frequency, while the ablation study shows limited sensitivity to the explicit capacity-state vector.

\end{document}